\documentclass[a4paper,aps,twocolumn,prl,10pt,intlimits]{revtex4-1}
\usepackage{amsmath,amsthm,amssymb,amsthm,dsfont,mathrsfs,bbm}
\usepackage{mathtools}
\usepackage[colorlinks]{hyperref}
\usepackage{multirow}
\usepackage[usenames,dvipsnames,svgnames,table]{xcolor}
\usepackage{asymptote}
\usepackage[caption=false]{subfig}
\usepackage[english]{babel}

\newcommand{\Bb}{\mathbf{b}}
\newcommand{\Br}{\mathbf{r}}
\newcommand{\Bmu}{\boldsymbol{\mu}}

\newcommand{\mB}{\mathcal{B}}
\newcommand{\mR}{\mathcal{R}}

\newcommand{\R}{\mathds{R}}
\newcommand{\Z}{\mathds{Z}}

\DeclareMathOperator{\dd}{d}

\DeclareMathOperator{\e}{e}
\newcommand{\abs}[1]{{\left|#1\right|}}
\newcommand{\norm}[1]{{\left\|#1\right\|}}

\begin{document}
\title{Quadratic stochastic Euclidean bipartite matching problem}
\author{Sergio Caracciolo}\email{sergio.caracciolo@mi.infn.it}
\affiliation{Dipartimento di Fisica, University of Milan and INFN, via Celoria 16, I-20133 Milan, Italy}
\author{Gabriele Sicuro}\email{sicuro@cbpf.br}\email{gabriele.sicuro@for.unipi.it}
\affiliation{Centro Brasileiro de Pesquisas F\'isicas, Rua Xavier Sigaud 150, 22290--180 Rio de Janeiro -- RJ, Brazil}

\date{\today}
\begin{abstract}
We propose a new approach for the study of the quadratic stochastic Euclidean bipartite matching problem between two sets of $N$ points each, $N\gg 1$. The points are supposed independently randomly generated on a domain $\Omega\subset\R^d$ with a given distribution $\rho(\mathbf x)$ on $\Omega$. In particular, we derive a general expression for the correlation function and for the average optimal cost of the optimal matching. A previous ansatz for the matching problem on the flat hypertorus is obtained as particular case. 
\end{abstract}
\maketitle

The \textit{Euclidean bipartite matching problem} (\textsc{Ebmp}) was firstly introduced and studied by \textcite{monge} in 1781. It is an \textit{assignment problem} in which an underlying geometric structure is present. Assignment problems are of paramount importance in theoretical computer science \cite{plummer1986,Papadimitriou2013} and a polynomial-time algorithm, the celebrated \textit{Hungarian algorithm} \cite{Kuhn,Munkres1957,Edmonds1972}, is available for their solution. In the \textsc{Ebmp} two sets of $N$ points, let us call them $\mR\coloneqq\{\Br_i\}_{i=1,\dots,N}$ and $\mB\coloneqq\{\Bb_i\}_{i=1,\dots,N}$, are considered on a domain $\Omega\subseteq\R^d$ in $d$ dimensions. The problem, in its \textit{quadratic} version, asks for the permutation $\pi\in\mathcal S_N$, $\mathcal S_N$ symmetric group of $N$ elements, such that the cost functional
\begin{equation}\label{costo}
E_N[\pi]\coloneqq\frac{1}{N}\sum_{i=1}^N\left\|\Bmu_{\pi(i)}(i)\right\|^2
\end{equation}
is minimized. In the previous formula we have introduced
\begin{equation}
\Bmu_{j}(i)\coloneqq\Bb_{j}-\Br_i
\end{equation}
and we have denoted by $\left\|\bullet\right\|$ the Euclidean norm in $\R^d$.

Matching problems appear in many different physical, biological and computational applications. The (linear) \textsc{Ebmp}, for example, was introduced by Monge to optimize the transport cost of soil from $N$ mining sites to $N$ construction sites. The problem of covering a given lattice with dimers can also be reformulated as a matching problem \cite{Baxter2007}, whereas, in computational biology, matching techniques are applied to pattern recognition problems \cite{Gusfield1997}. In computer vision, the quadratic \textsc{Ebmp} is at the basis of many image stitching and stereographic reconstruction algorithms \cite{szeliski2010}. Finally, the quadratic cost functional in Eq.~\eqref{costo} plays a special role in physical applications. Indeed, it was used by \textcite{Tanaka1978} in the study of Boltzmann equation, and by \textcite{Brenier1999} in his variational formulation of Euler incompressible fluids.

In many applications, however, the parameters (for example, the positions of the points) are affected by uncertainty, and the matching problem is a \textit{stochastic} (or random) optimization problem. In this case, the \textit{average} properties of the solution are of some interest. 

Many analytical and numerical techniques, derived from statistical physics \cite{Altarelli2011,mezard2009information}, were successfully applied to the study of stochastic optimization problems. In particular, in the \textit{random assignment problem} (\textsc{rap}), the quantities $\|\boldsymbol\mu_j(i)\|$ are considered independent and identically distributed random variables and the Euclidean structure is completely neglected. The \textsc{rap} was one of the first stochastic optimization problems to be solved using the theory of disordered systems by \textcite{Mezard1985}. Their results, obtained for $N\to\infty$, were rigorously derived years later by \textcite{Aldous2001}. Subsequently, \textcite{linusson2004} and \textcite{nair2005}, in two remarkable papers, proved independently Parisi's conjecture \cite{parisi1998} about the average optimal cost at finite $N$. They were able to prove also the more general Coppersmith--Sorkin conjecture \cite{coppersmith1999}, regarding the average optimal cost in the so-called \textit{$k$-assignment problem}.

In the present Letter we deal with the stochastic \textsc{Ebmp} (\textsc{sEbmp}). In the \textsc{sEbmp} the two sets of $N$ elements, $\mR$ and $\mB$ respectively (the \textit{instance} of the problem), are obtained extracting $2N$ points independently with a given probability distribution density $\rho(\mathbf x)$ on the domain $\Omega$. We are interested in the average properties of the optimal matching, and in particular in the optimal matching cost and in correlation functions. In contrast with the \textsc{rap}, in our case an Euclidean correlation appears among different values $\|\Bmu_{j}(i)\|$. This correlation is due to the underlying geometric structure. Denoting by $\overline\bullet$ the average over all instances, the \textit{average optimal cost} (\textsc{aoc}) is 
\begin{equation}
E_N\coloneqq \overline{\min_\pi E_N[\pi]}.
\end{equation}
This problem was studied perturbatively, under the assumptions $\rho(\mathbf x)=1$ and $\Omega\equiv[0,1]^d$, by \textcite{Mezard1988}, using the \textsc{rap} as a \textit{mean field approximation}. Their predictions were later confirmed numerically \cite{Boutet1997,Houdayer1998}. In \cite{Caracciolo2014} a proper scaling ansatz was adopted to evaluate directly the \textsc{aoc} and its finite size correction in any dimension, assuming a uniform distribution on the hypertorus. The one dimensional problem, again under the hypothesis of uniform distribution, was exactly solved in \cite{Boniolo2012,Caracciolo2014b}. 

Inspired by the celebrated Monge--Kantorovi\v c theory of optimal transportation \cite{villani2008optimal,Ambrosio2003}, we propose here a very general framework for the solution of the problem. Under the hypothesis that the points are generated using \textit{the same probability distribution density}, we can indeed write down a quadratic functional in the large $N$ limit. This functional can be used to compute every correlation function of the optimal solution of the quadratic \textsc{sEbmp} and to evaluate the scaling of the \textsc{aoc}.

Let us firstly consider a bounded $d$--dimensional domain $\Omega\subset\R^d$. Let $\mR\coloneqq\{\Br_i\}_{i=1,\dots,N}$ and $\mB\coloneqq\{\Bb_i\}_{i=1,\dots,N}$, be two sets, each one consisting of $N$ points independently generated with the same probability distribution density $\rho(\mathbf x)>0$ on $\Omega\setminus\partial\Omega$, $\partial\Omega$ boundary of $\Omega$. We introduce the following empirical measures
\begin{subequations}
\begin{align}
\rho_\mR(\mathbf x)&\coloneqq\frac{1}{N}\sum_{i=1}^N\delta^{(d)}\left(\mathbf x-\Br_i\right),\\\rho_\mB(\mathbf x)&\coloneqq\frac{1}{N}\sum_{i=1}^N\delta^{(d)}\left(\mathbf x-\Bb_i\right).
\end{align}\end{subequations}
We define also the functional
\begin{equation}
E[\Bmu]\coloneqq\int_{\Omega}\norm{\Bmu(\mathbf x)}^2\rho_\mR(\mathbf x)\,d\mathbf  x
\end{equation}
for a map $\Bmu\colon\Omega\to\R^d$. The previous functional provides a correct matching cost, Eq.~\eqref{costo}, if, and only if, $\Bmu(\Br_i)=\Bb_{\pi(i)}-\Br_i$ for a certain permutation $\pi\in\mathcal S_N$. This additional constraint implies
\begin{equation}
\int_{\Omega}\delta^{(d)}\left(\mathbf x-\mathbf y-\Bmu(\mathbf y)\right)\rho_\mR(\mathbf y)\,d\mathbf  y=\rho_\mB(\mathbf x).\label{const}
\end{equation}
We can write down a ``partition function'' for our problem introducing a proper Lagrange multiplier $\varphi(\mathbf x)$ to impose the constraint in Eq.~\eqref{const},
\begin{equation}
Z(\beta)\propto\int\left[\mathcal D\Bmu\right]\int_{-i\infty}^{+i\infty}\left[\mathcal D\varphi\right]\e^{-\beta S[\Bmu,\varphi]},
\end{equation}
the optimal solution being recovered for $\beta\to+\infty$. The exponent in the functional integral is
\begin{multline}
S[\Bmu,\varphi]\coloneqq\frac{1}{2}E[\Bmu]\\+\int_{\Omega}\left[\varphi(\mathbf x)\rho_\mB(\mathbf x)-\varphi(\mathbf x+\Bmu(\mathbf x))\rho_\mR(\mathbf x)\right]\,d\mathbf x\\
=-\int_{\Omega}\left[\varphi(\mathbf x)\varrho(\mathbf x)+\rho_\mR(\mathbf x)\Bmu(\mathbf x)\cdot\nabla\varphi(\mathbf x)\right]\,d\mathbf x\\
+ \frac{1}{2}E[\Bmu]+s[\Bmu,\varphi],\end{multline}
where $s[\Bmu,\varphi]=O\left(\norm{\Bmu}^2\varphi\right)$ are higher order nonlinear terms in the fields obtained from the Taylor series expansion of $\varphi(\mathbf x+\Bmu)$ around $\Bmu=\mathbf 0$. We introduced also
\begin{equation}
\varrho(\mathbf x)\coloneqq \rho_\mR(\mathbf x)-\rho_\mB(\mathbf x).
\end{equation}
Observing that $\rho_\mR(\mathbf x)$ is almost surely zero everywhere on the boundary, the Euler--Lagrange equations are 
\begin{subequations}
\begin{align}
\varrho(\mathbf x)&=\nabla\cdot\left(\rho_\mR(\mathbf x)\Bmu(\mathbf x)\right)-\frac{\delta s[\Bmu,\varphi]}{\delta\varphi(\mathbf x)},\label{el1}\\
\rho_\mR(\mathbf x)\Bmu(\mathbf x)&=\rho_\mR(\mathbf x)\nabla\varphi(\mathbf x)-\frac{\delta s[\Bmu,\varphi]}{\delta\Bmu(\mathbf x)}.\label{el2}
\end{align}\end{subequations}

It is well known that in the $N\to\infty$ limit, the empirical measures $\rho_\mR(\mathbf x)$ and $\rho_\mB(\mathbf x)$ both converge (in weak sense) to $\rho(\mathbf x)$. In this limit the optimal field $\Bmu^*$ is trivially $\Bmu^*(\mathbf x)\equiv \mathbf 0$ $\forall\mathbf x\in\Omega$. For $N\gg 1$ we expect that the relevant contribution is given by small values of $\norm{\Bmu}$ and the nonlinear terms in $s$ are higher order corrections to the leading quadratic terms. 
The saddle point equations simplify as
\begin{subequations}
\begin{align}
\varrho(\mathbf x)&=\nabla\cdot\left(\rho(\mathbf x)\Bmu(\mathbf x)\right),\label{elq1}\\
\Bmu(\mathbf x)&=\nabla\varphi(\mathbf x).\label{elq2}
\end{align}\end{subequations}
The strict analogy between our problem and an electrostatic problem is evident. The field $\Bmu$ plays the role of an electric field $\mathbf E$,  $-\varphi$ is the scalar potential, and, indeed, it acts as a Lagrange multiplier which implements the Gauss law, whereas $\rho$ corresponds to a dielectric function $\epsilon$ in a linear dielectric medium, in such a way that the equivalent of the displacement field $\mathbf D = \epsilon\mathbf E$  is  $\rho\nabla\varphi$. The $\mathcal{B}$-points and the $\mathcal{R}$-points play the role of point-like charges of opposite sign, being the overall charge $\int_\Omega\varrho(\mathbf x)\,d\mathbf x=0$.
It is remarkable that Eq.~\eqref{elq2} reproduces the known result in measure theory that the transport field is a gradient \cite{villani2008optimal} but, in our approach, this is specified as the gradient of the introduced Lagrange multiplier. We impose Neumann boundary conditions
\begin{equation}\label{Nbc}
\left.\nabla_{\mathbf n(\mathbf x)}\varphi(\mathbf x)\right|_{\mathbf x\in\partial\Omega}\equiv \left.\nabla\varphi(\mathbf x)\cdot\mathbf n(\mathbf x)\right|_{\mathbf x\in\partial\Omega}=0,
\end{equation}
where $\mathbf n(\mathbf x)$ is the normal unit vector to the boundary in $\mathbf x\in\partial\Omega$. This condition guarantees that the shape of the boundary is not modified in the $N\to\infty$ limit. We can therefore compute $\varphi$ as the solution of the following equation on $\Omega$ with the given boundary conditions
\begin{equation}\label{eqphi}
\nabla\cdot\left[\rho(\mathbf x)\nabla\varphi(\mathbf x)\right]=\varrho(\mathbf x).
\end{equation}
To solve Eq.~\eqref{eqphi}, we use the \textit{modified} Green's function $G_\rho(\mathbf x,\mathbf y)$ of the operator $\nabla\cdot\left[\rho(\mathbf x)\nabla\bullet\right]$ on $\Omega$, defined by
\begin{multline}\label{green}
\nabla_\mathbf{x}\cdot\left[\rho(\mathbf x)\nabla_\mathbf{x}G_\rho(\mathbf x,\mathbf y)\right]=\delta^{(d)}\left(\mathbf x-\mathbf y\right)-\frac{1}{\abs\Omega},\\\text{with }\left.\frac{\partial G_\rho(\mathbf x,\mathbf y)}{\partial\mathbf n(\mathbf x)}\right|_{\mathbf x\in\partial\Omega}=0.
\end{multline}
In Eq.~\eqref{green}, $\abs\Omega$ is the Lebesgue measure of $\Omega$.
We can write an explicit expression for $\Bmu(\mathbf x)$ as
\begin{equation}
\Bmu(\mathbf x)=\int_{\Omega}\nabla_\mathbf{x}G_\rho(\mathbf x,\mathbf y)\varrho(\mathbf y)\,d\mathbf  y.
\end{equation}
Averaging over the disorder, we easily obtain the following {two-point correlation function}
\begin{widetext}\begin{multline}\label{correlazione}
C(\mathbf x,\mathbf y)\coloneqq\overline{\Bmu(\mathbf x)\cdot\Bmu(\mathbf y)}
=\iint_{\Omega_N(\mathbf x)\times \Omega_N(\mathbf y)}\left[\overline{\varrho(\mathbf z)\varrho(\mathbf w)}\nabla_\mathbf{x}G_\rho(\mathbf x,\mathbf z)\cdot\nabla_\mathbf{y}G_\rho(\mathbf y,\mathbf w)\right]\,d\mathbf z\,d\mathbf w\\
=\frac{2}{N}\int_{\Omega_N(\mathbf x,\mathbf y)}\left[\rho(\mathbf z)\nabla_\mathbf{x}G_\rho(\mathbf x,\mathbf z)\cdot\nabla_\mathbf{y}G_\rho(\mathbf y,\mathbf z)\right]\,d\mathbf z
-\frac{2}{N}\iint_{\Omega_N(\mathbf x)\times \Omega_N(\mathbf y)}\left[\rho(\mathbf z)\rho(\mathbf w)\nabla_\mathbf{x}G_\rho(\mathbf x,\mathbf z)\cdot\nabla_\mathbf{y}G_\rho(\mathbf y,\mathbf w)\right]\,d\mathbf z\,d\mathbf w,
\end{multline}\end{widetext}
where we denoted by $\overline{\bullet}$ the average over all instances. In the previous equation we used the following result
\begin{equation}\label{overrho}
\overline{\varrho(\mathbf z)\varrho(\mathbf w)}=2\frac{\rho(\mathbf z)}{N}\left[\delta^{(d)}\left(\mathbf z-\mathbf w\right)-\rho(\mathbf w)\right].
\end{equation}
Moreover, we introduced \textit{a proper cutoff} to avoid divergences in the expression $NC(\mathbf x,\mathbf y)$ and take into account finite size effects. This cutoff has indeed an intuitive explanation. Let $\delta_N$ be the scaling law in $N$ of the average distance between two \textit{nearest neighbor} points randomly generated on $\Omega$ accordingly to $\rho(\mathbf x)$. We introduced
\begin{subequations}
\begin{align}
\Omega_{N}(\mathbf x)\coloneqq&\{\mathbf y\in\Omega\colon \norm{\mathbf x-\mathbf y}>\alpha \delta_N\},\\
\begin{split}\Omega_{N}(\mathbf x,\mathbf y)\coloneqq&\{\mathbf z\in\Omega\colon \norm{\mathbf x-\mathbf z}>\alpha \delta_N\\&\text{ and }\norm{\mathbf y-\mathbf z}>\alpha \delta_N\},\quad \alpha\in \R^+.\end{split}
\end{align}\end{subequations}
Observe that $\delta_N\xrightarrow{N\to\infty}0$. The scaling quantity $\delta_N$ takes into account the nonzero characteristic length for finite $N$. The results of the computation may depend upon the regularizing parameter $\alpha$. 

Eq.~\eqref{correlazione} provides a recipe for the calculation of the \textsc{aoc} and for the correlation function in the \textsc{sEbmp}. In particular, in our approximation we have that
\begin{equation}\label{ene} E_N\simeq \int_\Omega C(\mathbf x,\mathbf x)\rho(\mathbf x)\,d\mathbf  x.\end{equation}
If no regularization is required ($\alpha=0$) we can write
\begin{equation}
E_N\simeq \frac{2}{N}\iint_{\Omega\times\Omega} \rho(\mathbf x)\left[\rho(\mathbf y)G_\rho(\mathbf x,\mathbf y)-\frac{G_\rho(\mathbf x,\mathbf x)}{\abs{\Omega}}\right]\,d\mathbf  x\,d\mathbf  y.
\end{equation}

Let us now consider the one dimensional problem, $\Omega=[a,b]\subset\R$, and a certain probability density distribution $\rho(x)$ on $\Omega$. In this case we can explicitly write ($\alpha=0$) the correlation function and the \textsc{aoc}, from Eq.~\eqref{correlazione} and Eq.~\eqref{ene} respectively. Imposing Neumann boundary conditions $\left.\partial_x\varphi(x)\right|_{x=a}=\left.\partial_x\varphi(x)\right|_{x=b}=0$
\begin{subequations}\label{correneline}
\begin{align}\label{corrline}
C(x,y)&=\frac{2}{N}\frac{\Phi_\rho(\min\{x,y\})-\Phi_\rho(x)\Phi_\rho(y)}{\rho(x)\rho(y)},\\
\label{eneline}
E_N&=\frac{2}{N}\int_{a}^{b}\frac{\Phi_\rho(x)(1-\Phi_\rho(x))}{\rho(x)}\dd x,
\end{align}
\end{subequations}
where we introduced the cumulative function
\begin{equation}
\Phi_\rho(x)\coloneqq\int_{a}^x\rho(\xi)\dd \xi.
\end{equation}

Our approach is suitable for many applications. In the following we shall provide some examples and numerical verifications. 

\paragraph{Matching problem on the interval} 
\begin{figure}
\centering
\subfloat[Correlation function $C(x,x)$ and $C(x,-x)$ for $N=3000$, obtained averaging over $5000$ instances of the problem. We compare with the theoretical predictions obtained from Eq.~\eqref{corrline}.]{\includegraphics[width=\columnwidth]{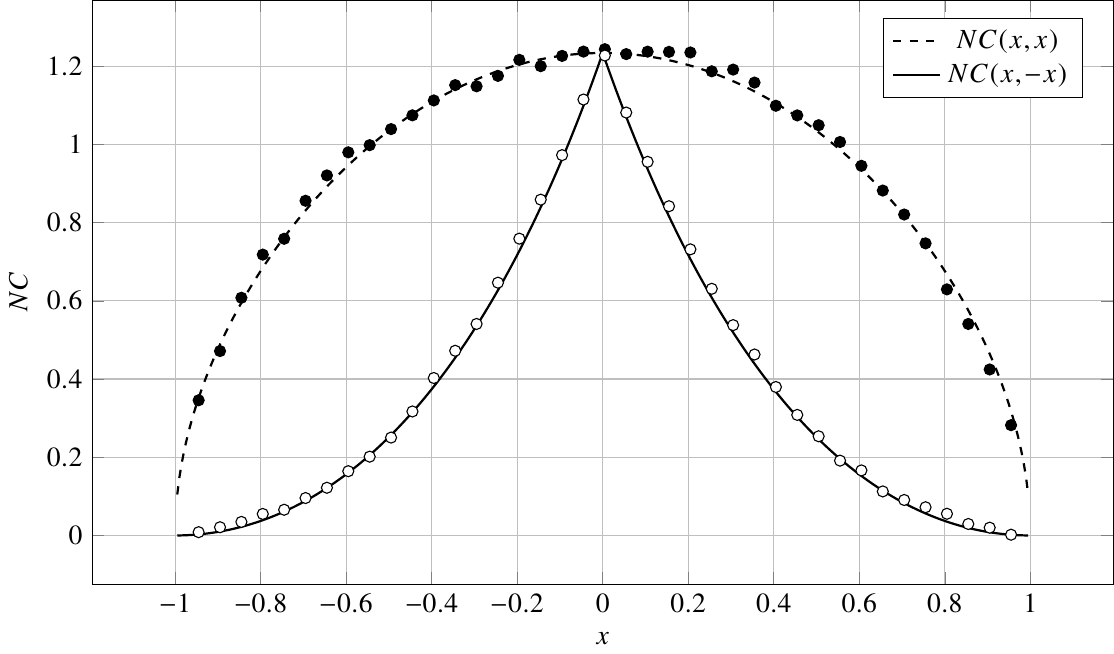}}\\
\subfloat[\textsc{Aoc} obtained averaging over $5000$ instances. We compare with the theoretical prediction obtained from Eq.~\eqref{eneline}, presented in Eq.~\eqref{enecircle}.]{\includegraphics[width=\columnwidth]{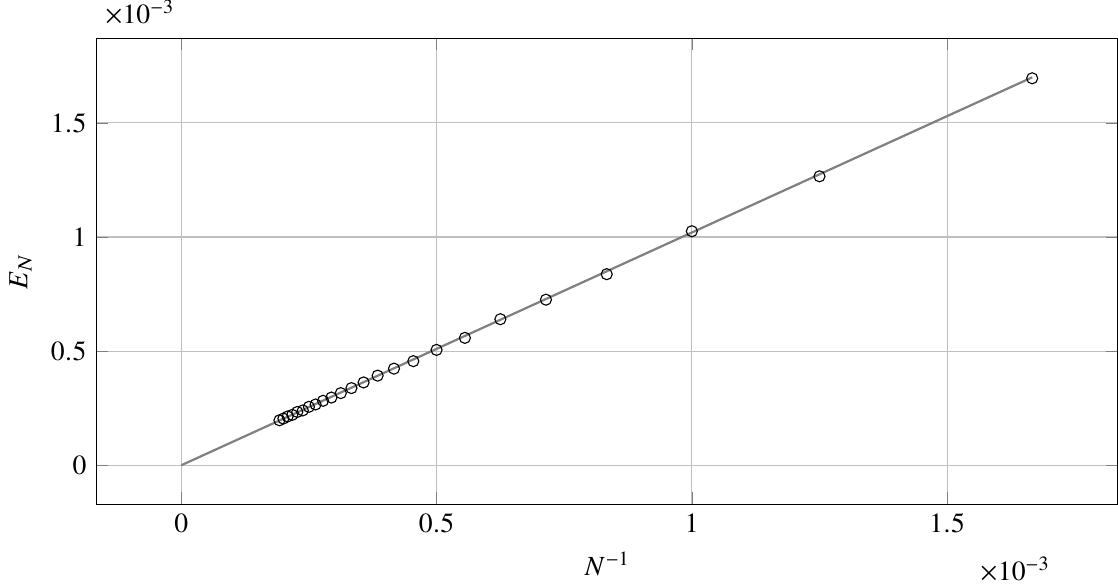}}
\caption{\textsc{sEbmp} on the real line with points generated using a semi-circle distribution, Eqs.~\eqref{circle}.\label{Fcircle}}
\end{figure}
As application of Eqs.~\eqref{correneline}, let us assume, for example, a \textit{semicircle distribution} on $\Omega\equiv[-1,1]$,
\begin{subequations}\label{circle}
\begin{align}
\rho(x)&=\frac{2\sqrt{1-x^2}}{\pi}\qquad x\in[-1,1],\\
\Phi_\rho(x)&= 1+\frac{x\sqrt{1-x^2}-\arccos x}{\pi}.
\end{align}\end{subequations}
We can compute straightforwardly the correlation function and the \textsc{aoc} using Eqs. \eqref{correneline}. In particular, we obtain the non-trivial result
\begin{equation}\label{enecircle}
E_N=\frac{1}{N}\left(\frac{\pi^2}{6}-\frac{5}{8}\right)+o\left(\frac{1}{N}\right).
\end{equation}
In Fig.~\ref{Fcircle} we compare the numerical results with the analytical predictions, showing the excellent agreement both for the correlation function and for the \textsc{aoc}.

Observe also that eq.~\eqref{corrline} provides the correct correlation function for the \textsc{sEbmp} on $\Omega\equiv [0,1]$ with uniform distribution. Assuming indeed $\rho(x)=\theta(x)\theta(1-x)$, being $\theta(x)$ the Heaviside function, we have
\begin{subequations}
\begin{eqnarray}
C(x,y)&=&\begin{cases}2\tfrac{\min\{x,y\}-xy}{N}&(x,y)\in[0,1]^2\\0&\text{otherwise},\end{cases}\\
E_N& =&\frac{1}{3N}+o\left(\frac{1}{N}\right).
\end{eqnarray}
\end{subequations}
Similar expressions have been derived, using a different approach, in \cite{Boniolo2012,Caracciolo2014b} for the grid--Poisson matching problem.

\paragraph{Matching problem on the unit square}
\begin{figure}{\includegraphics[width=\columnwidth]{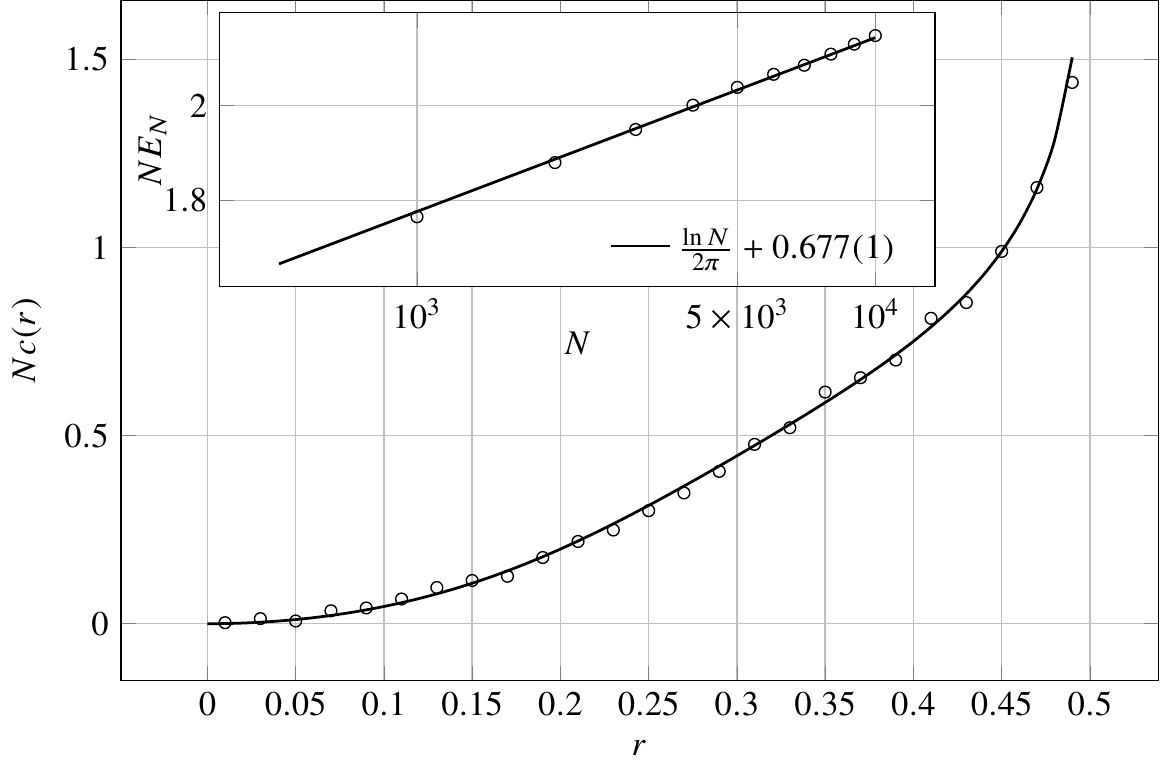}}
  \caption{Matching problem on the square. In the main plot, correlation function between points on the diagonals of the square, see Eq.~\eqref{crsq}, obtained for $N=3000$ and averaging over $2\cdot 10^4$ instances. We compare with our analytical prediction. In the smaller plot, we compare our theoretical prediction for the \textsc{aoc}, Eq.~\eqref{CostoQuadrato} with numerical results obtained averaging over $2\cdot 10^4$ instances. In particular, the value of $\gamma=0.677(1)$ is obtained by a fit procedure.}\label{FQ}
\end{figure}
Let us now consider $\Omega=\{(x_1,x_2)\in\R^2\colon 0<x_1<1,\ 0<x_2<1\}\subset\R^2$. Let us suppose also that $\rho(\mathbf x)=1$ on $\Omega$. Using Eq.~\eqref{correlazione} we can compute $C(\mathbf x,\mathbf y)$ as function of the modified Green's function of the Laplacian on the square with Neumann boundary condition $G_s(\mathbf x,\mathbf y)$. However, it can be seen that $NC(\mathbf x,\mathbf x)\to\infty$ for $N\to\infty$ and we need to impose a regularizing cutoff to properly evaluate this quantity. Being in this case $\delta_N\sim \frac{1}{\sqrt{N}}$, a regularization procedure gives us \footnote{For the details, see the Supplemental material.}
\begin{equation}
E_N=\frac{\ln N}{2\pi N}+\frac{\gamma}{N} +o\left(\frac{1}{N}\right),\label{CostoQuadrato}
\end{equation}
for some constant $\gamma$. Note that the leading term is exactly the same obtained for the \textsc{sEbmp} on the $2$-dimensional torus \cite{Caracciolo2014}. In Fig.~\ref{FQ} we plotted the numerical results for the \textsc{aoc} and we compare with the previous prediction. Moreover, we compare also our numerical results with the theoretical prediction for $c(r)$, defined as the correlation function between points on the diagonals of the square as follows
\begin{equation}
\mathbf x_r\coloneqq(r,r),\ \mathbf y_r\coloneqq(r,1-r),\quad c(r)\coloneqq NC(\mathbf x_r,\mathbf y_r).\label{crsq}
\end{equation}

\paragraph{Matching problem on the flat hypertorus} Finally, we consider the domain $\Omega\equiv[0,1]^d\subset\R^d$ with \textit{periodic boundary conditions}, i.e., we deal with the \textsc{sEbmp} on the flat unit hypertorus in $d$ dimensions $\mathsf T^d\coloneqq\R^d/\Z^d$. We can restate the results above for this case simply by substituting the Neumann boundary conditions in \eqref{eqphi} and \eqref{green} with {periodic} boundary conditions. Moreover, the Euclidean distance in \eqref{costo} between the points $\mathbf x=(x_i)_{i=1,\dots,d}$ and $\mathbf y=(y_i)_{i=1,\dots,d}$ in $\Omega$ must be intended as
\begin{equation}
\norm{\mathbf x-\mathbf y}^2\coloneqq\sum_{i=1}^d\left[\min\left(\abs{x_i-y_i},1-\abs{x_i-y_i}\right)\right]^2.
\end{equation} 
Assuming $\rho(\mathbf x)=1$ and $\alpha=0$, then $G_\rho(\mathbf x,\mathbf y)\equiv G_d(\mathbf x-\mathbf y)$, where $G_d(\mathbf x-\mathbf y)$ is the Green's function of the Laplacian on the unit flat hypertorus $\mathsf T^d$
\begin{equation}\label{green0}
\nabla_\mathbf{x}^2G_d(\mathbf x-\mathbf y)=\delta^{(d)}\left(\mathbf x-\mathbf y\right)-1.
\end{equation}
Under these hypotheses we have that, up to higher order terms, we can formally write
\begin{equation}\label{ansatz}
C(\mathbf x,\mathbf y)=-\frac{2}{N}G_d(\mathbf x-\mathbf y),\quad E_N=-\frac{2}{N}G_d(\mathbf 0).
\end{equation}
For $d=1$ Eqs.~\eqref{ansatz} have the form
\begin{equation}\label{cfcosto1d}
C(x,y)=\frac{1-6\abs{x-y}\left(1-\abs{x-y}\right)}{6 N},\quad E_N=\frac{1}{6N}+o\left(\frac{1}{N}\right).
\end{equation}

Eqs.~\eqref{ansatz} were adopted as a working ansatz in \cite{Caracciolo2014,Caracciolo2015} and they were used to derive both the scaling of the \textsc{aoc} and the correlation functions of the \textsc{sEbmp} on $\mathsf T^d$. For $d\geq 2$, however, $G_d(\mathbf 0)$ is a \textit{divergent} quantity. In this case, a nonzero value of $\alpha$ must be taken and the regularization must be performed, as shown in \cite{Caracciolo2014}.

\paragraph{Conclusions} The presented approach allows us to go beyond the mean field approximation in the \textsc{sEbmp}, and to easily evaluate the scaling behavior of the \textsc{aoc} and other useful quantities, like the correlation functions of the optimal solution. A deep connection is established among the theory of combinatorial optimization, the theory of optimal transport and the theory of disordered systems and stochastic processes. Indeed, even if optimal transport theory has been already successfully applied to many different physical problems (kinetic theory, fluidodynamics\dots), the study of the properties of the solution in presence of disorder (e.g., uncertainty on the distribution parameters) is a highly nontrivial task. This interesting research line is still largely unexplored by both physicists and mathematicians and we hope that these results will allow further studies in this direction. Finally, the method presented here may be useful in the analysis of other stochastic Euclidean optimization problems, where both disorder and geometric constraints appear.

\begin{acknowledgments}The authors are grateful to Luigi Ambrosio, Carlo Lucibello, Giorgio Parisi and Andrea Sportiello for fruitful discussions. G.S.\@ acknowledges the financial support of the John Templeton Foundation and the computational facilities provided by the INCT-SC.
\end{acknowledgments}

\appendix
\section*{Supplemental material}
\subsection*{Evaluation of the correlation function for the \textsc{sEbmp} on the unit square}
We derive here the correct scaling of the \textsc{aoc} for the \textsc{sEbmp} on the square $\Omega=\{\mathbf x=(x_1,x_2)\in\R^2\colon 0<x_1<1\text{ and }0<x_2<1\}$ with $\rho(\mathbf x)=1$. In the following $\mathbf x=(x_1,x_2)$ and $\mathbf y=(y_1,y_2)$. Following the general recipe, the modified Green's function on the square with Neumann conditions is
\begin{equation}
\nabla_{\mathbf{y}}^2G_s(\mathbf x,\mathbf y)=\delta^{(d)}\left(\mathbf x-\mathbf y\right)-1,\quad \left.\frac{\partial G_s(\mathbf x,\mathbf y)}{\partial\mathbf n(\mathbf y)}\right|_{\mathbf y\in\partial\Omega}=0.
\end{equation}
\begin{widetext}
The previous problem can be solved using a mode expansion. We find
\begin{multline}
G_s(\mathbf x,\mathbf y)=4\sum_{n_1=1}^\infty\sum_{n_2=1}^\infty \frac{\cos(n_1\pi x_1)\cos(n_2\pi x_2)\cos(n_1\pi y_1)\cos(n_2\pi y_2)}{\pi^2(n_1^2+n_2^2)}\\
+2\sum_{n=1}^\infty\frac{\cos(n\pi x_1)\cos(n\pi y_1)+\cos(n\pi x_2)\cos(n\pi y_2)}{\pi^2 n^2}.
\end{multline}
Observing that $\int_\Omega G_s(\mathbf x,\mathbf y)\dd^2 y=0$, the correlation function on the square can be expressed as ($\alpha\equiv 0$)
\begin{multline}\label{Cf}
C(\mathbf x,\mathbf y)=\frac{2}{N}\left(\frac{\partial^2}{\partial x_1\partial y_1}+\frac{\partial^2}{\partial x_2\partial y_2}\right)\int_\Omega G_s(\mathbf x,\mathbf z)G_s(\mathbf y,\mathbf z)\,d\mathbf  z=\\
=\frac{8}{N\pi^2}\sum_{n_1=1}^\infty\sum_{n_2=1}^\infty \frac{n_1^2\sin(n_1\pi x_1)\sin(n_1\pi y_1)\cos(n_2\pi x_1)\cos(n_2\pi y_2)+n_2^2\cos(n_1\pi x_1)\cos(n_1\pi y_1)\sin(n_2\pi x_1)\sin(n_2\pi y_2)}{(n_1^2+n_2^2)^2}\\
+\frac{4}{N\pi^2}\sum_{n=1}^\infty\frac{\sin(n\pi x_1)\sin(n\pi y_1)+\sin(n\pi x_2)\sin(n\pi y_2)}{n^2}.
\end{multline}
The quantity $NC(\mathbf x,\mathbf y)$ is finite everywhere, except for $\mathbf x\equiv\mathbf y$.

In particular, if $\mathbf x\equiv \mathbf x_r=(r,r)$ and $\mathbf y\equiv \mathbf y_r=(r,1-r)$, $r\in[0,1]$, we obtain:
\begin{multline}
c(r)\coloneqq C(\mathbf x_r,\mathbf y_r)=
4\sum_{n=1}^\infty\frac{\sin^2((2n-1)\pi r)}{\pi (2n-1)^2}\\
+\frac{8}{N\pi^2}\sum_{n_1=1}^\infty\sum_{n_2=1}^\infty (-1)^{n_2}\frac{n_1^2\sin^2(n_1\pi r)\cos^2(n_2\pi r)-n_2^2\cos^2(n_1\pi r)\sin^2(n_2\pi r)}{(n_1^2+n_2^2)^2}.
\end{multline}
\end{widetext}

\begin{figure}\includegraphics[width=0.7\columnwidth]{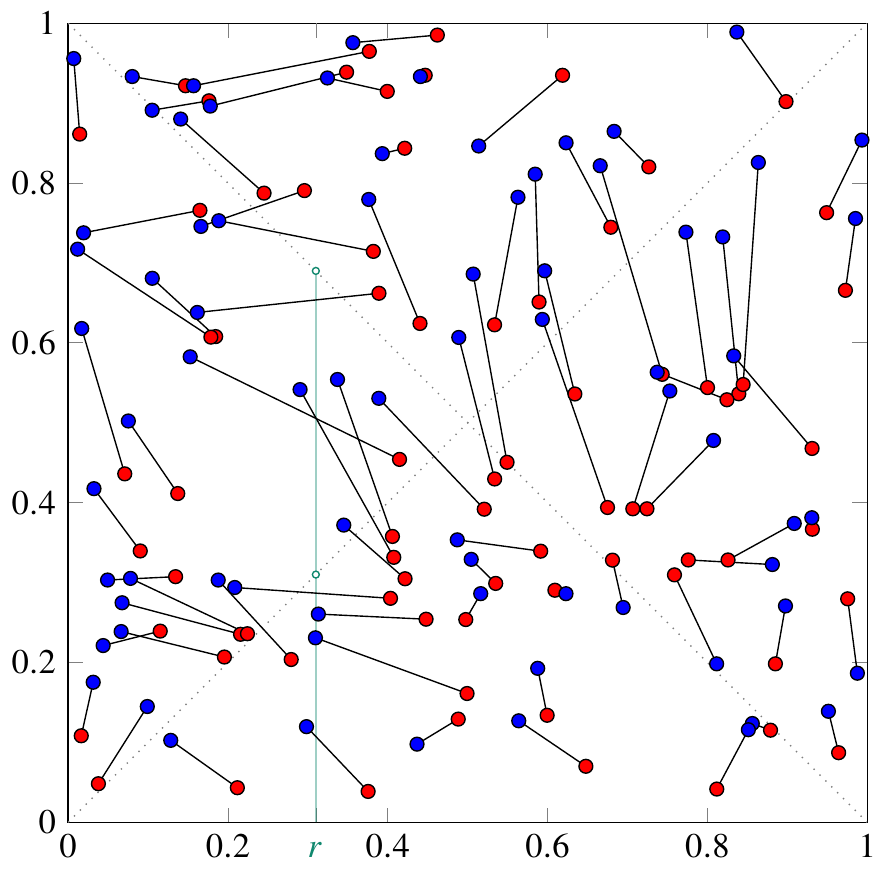}\caption{(Color online) \textsc{Ebmp} on the square. The geometrical meaning of the variable $r$ in Eq.~\eqref{crsq} is also depicted.}\label{FM}\end{figure}
Finally, let us compute the \textsc{aoc}. As anticipated, $C(\mathbf x,\mathbf x)$ is a divergent quantity and therefore a proper cutoff must be introduced. In particular, note that the average distance between two points uniformly randomly generated on $\Omega$ scales as $\delta_N=\frac{1}{\sqrt N}$. It follows that the cutoff in the Fourier space can be imposed requiring $n_1^2+n_2^2<\alpha N$ and $n^2<\tilde \alpha N$ in the sums appearing in \eqref{Cf}. Here $\tilde \alpha$ is a regularizing parameter. Therefore
\begin{equation}
\int_\Omega C(\mathbf x,\mathbf x)\rho(\mathbf x)\dd^2x=\frac{1}{2\pi^2N}\sum_{\substack{\mathbf n\in\Z^2\\0<\norm{\mathbf n}^2<\tilde \alpha N}}\frac{1}{\norm{\mathbf n}^2}+O\left(\frac{1}{N}\right).
\end{equation}
The regularization of this sum is performed in \cite{Caracciolo2014}, obtaining
\begin{equation}
E_N=\frac{\ln N}{2\pi N}+\frac{\gamma}{N}+o\left(\frac{1}{N}\right),
\end{equation}
where $\gamma$ is a certain constant depending on the cutoff.
\bibliography{Biblio.bib}
\end{document}